\documentclass[pre,twocolumn,showpacs,floatfix]{revtex4}

\usepackage{graphicx}
\usepackage{dcolumn}
\usepackage{bm}

\begin{document}
\title{Diffusion and percolation in anisotropic random barrier models}
\author{Sebastian Bustingorry}
\affiliation{Centro At\'omico Bariloche,
8400 San Carlos de Bariloche, R\'{\i}o Negro, Argentina}

\date{\today}

\begin{abstract}
An anisotropic random barrier model is presented, in which the transition 
probabilities in different directions have different probability density 
functions.
At low temperatures, the anisotropic long--time diffusion coefficients, obtained
using an effective medium approximation, follow an Arrhenius temperature 
dependence, with the same activation
energy for each direction.
Such activation energy is related to the anisotropic 
percolation properties of the lattice, and can be analysed in terms of
the critical percolation path approximation.
The anisotropic effective medium approximation is shown to predict the correct 
percolation threshold for an anisotropic two--dimensional square lattice. In
addition, results are compared with numerical simulations using a fast 
kinetic Monte Carlo algorithm.
\end{abstract}

\pacs{05.40.-a,	
	05.60.Cd,	
	66.30.-h	
	}

\maketitle

\section{Introduction}

Diffusion in disordered media is an active field of research, due to its 
relevance in a wide variety of natural and industrial processes
\cite{haus87,bouc90,sahimiL}.
One of the traditional models for disorder is the random barrier model (RBM), 
which consists of equally energy minima separated by energy barriers, the height
of which
is randomly distributed according to a given probability density function (PDF).
In this model, a particle moves from one minimum to another by performing thermally activated 
jumps.

In these systems, diffusion properties can be studied either in time or 
frequency variables. Several studies have been conducted of diffusion 
properties in the isotropic RBM both under unbiased 
\cite{dyre94,dyreR,avra93,amba95,argy95,horn95} and biased 
\cite{arap97,avra98} conditions.
Following a frequency analysis (Refs. \cite{dyre94,dyreR} and references 
therein), the system may be characterized by a zero--frequency 
diffusion constant $D(s=0)$, and a characteristic 
frequency $s^*$, which marks the onset of frequency--dependent diffusion. 
$D(s=0)$ and $s^*$ follow Arrhenius laws with the same activation energy $E_c$.
Analogously, from a time variable standpoint, it takes a time 
$t^* \sim s^{*-1}$ for a particle to reach a long--time 
diffusion regime in the RBM, characterized by a diffusion constant 
$D(t \rightarrow \infty) \equiv D(\omega=0)$ \cite{argy95}. 
The activation energy $E_c$ depends on the percolation properties of the 
lattice and the PDF of the energy barriers.
This dependence is simply achieved by the critical percolation path 
approximation (CPPA), as shown for isotropic
problems \cite{ambe71,tyc89,ledo89}.

In view of the diversity of systems in 
which diffusion takes place, the anisotropic generalization of 
diffusion problems  has attracted considerable attention in the last years,
both under unbiased
\cite{bern74,parr87,tole92,reye00,cacerey,bust03,saad02} and biased 
\cite{bust00,bust02,huang02} conditions.
A few examples of anisotropic systems are porous reservoir rocks
\cite{sahimiL,saad02,koelman}, layered semiconducting compounds \cite{gallos},
and superconductor cuprates \cite{jhan96}.
When dealing with anisotropic conditions, diffusion properties are
independently studied in the different relevant directions of the system. 
It was recently shown that, for a 
two--dimensional anisotropic bond percolation model, different activation 
energies are found in each direction \cite{bust03}.

In the present paper, a two--dimensional unbiased diffusion process is 
studied with each direction characterized by a different continuous PDF.
The paper is organized as follows. In Sec. \ref{SII} the anisotropic RBM
is introduced. In Sec. \ref{SIII}, the model is studied within the anisotropic 
effective medium approximation (EMA). These results are 
compared against Monte Carlo simulations, whose numerical details are given in 
Sec. \ref{SIV}.
Section \ref{SV} is devoted to a description of the CPPA
ideas in anisotropic conditions, and in Sec. \ref{SVI} the concluding
remarks of the present paper are presented.

\section{Anisotropic Random Barrier Model\label{SII}}

Diffusion processes will be studied on a two--dimensional square lattice with 
static disorder.
Energy barriers are chosen from a given PDF
$\rho \left( E\right) $ at $t=0$ and are kept constant during the
diffusion process. Possible jumps are only allowed between nearest neighbors.
Once the energy barrier $E_{ij}$ between sites $i$ and $j$ is selected, the
transition rates $\omega _{ij}$ from site $i$ to site $j$ are determined
following an Arrhenius law 
\begin{equation}
\label{wdeE}
\omega _{ij}=\frac{\omega _0}{z}e^{-\beta E_{ij}},
\end{equation}
where $\omega _0$ is the constant jump rate, $z=4$ is the coordination number,
and $\beta =1/k_{B}T$ is the inverse temperature, with $k_B$ being the Boltzmann
constant.
The energy $E_{ij}$ characterizes the bond joining sites $i$ and $j$, therefore
$E_{ij}=E_{ji}$, and the forward $(i\rightarrow j)$ and backward 
$(j\rightarrow i)$ jumps have the same transition rate.

\begin{figure}[!tbp]
\includegraphics[angle=-90,width=8.5cm,clip=true]{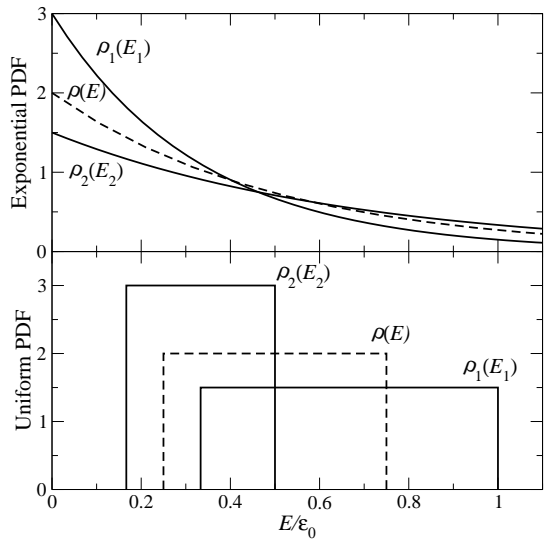}
\caption{\label{f1}
Exponential (upper panel) and uniform (lower panel) probability density 
functions. Solid lines represent the anisotropic cases ($\alpha =2$), 
and dashed lines the isotropic cases.}
\end{figure}

In order to introduce the anisotropic character of the system, the $E_{ij}$
energies are selected from different PDFs, depending on the orientation of
the bond joining sites $i$ and $j$. Let $1$ and $2$ be the main
directions of the square lattice, the key idea is to introduce $\rho _1(E_1)$
and $\rho _2(E_2)$ instead of a single PDF $\rho (E)$.
The model is characterized by anisotropy 
$\alpha =\epsilon _{1}/\epsilon _{2}$ and global mean
energy $\epsilon =(\epsilon _{1}+\epsilon _{2})/2$.
The mean energies in each direction, $\epsilon_1$ and $\epsilon_2$, are 
thus represented by $\epsilon_1=2\alpha \epsilon /(\alpha +1)$ and 
$\epsilon_2=2\epsilon /(\alpha+1)$.
In the present work, a constant value of $\epsilon =0.5 \epsilon_0$ is 
adopted, where $\epsilon_0$ sets the unit of energy, and the effects of having
$\alpha \neq 1$ are studied.
Two different anisotropic distributions will be considered: 
a) an exponential PDF 
\begin{eqnarray}
\label{pdfexp} 
\rho_1(E_1) =\frac{1}{\epsilon_1}e^{-E_1/\epsilon_1},
\qquad E_1\in \left[ 0,\infty \right), \nonumber \\*
\rho_2(E_2) =\frac{1}{\epsilon_2}e^{-E_2/\epsilon_2},
\qquad E_2\in \left[ 0,\infty \right),
\end{eqnarray}
and b) a uniform PDF
\begin{eqnarray}
\label{pdfuni}
\rho_1(E_1) =\frac{1}{2\delta_1 \epsilon_1},
\qquad E_1\in \left[(1-\delta_1)\epsilon_1,
(1+\delta_1)\epsilon_1\right], \nonumber \\* 
\rho_2(E_2) =\frac{1}{2\delta_2 \epsilon_2},
\qquad E_2\in \left[(1-\delta_2)\epsilon_2,(1+\delta_2)\epsilon_2\right],
\end{eqnarray}
where $\delta_1$ and $\delta_2$ serve to control different distribution widths 
in each direction.
This uniform PDF represents the most general
anisotropic extension of the isotropic uniform PDF used in Ref. \cite{argy95} to
study diffusion in RBM. 
In the following, and for the sake of simplicity, the widths of the uniform PDF
will be $\delta_1 = \delta_2 = 0.5$.
Figure \ref{f1} shows the exponential and uniform PDFs for $\alpha =1$ and 
$\alpha =2$.

\section{Anisotropic EMA: Low temperature predictions\label{SIII}}

The EMA self--consistent conditions provide a method for obtaining the diffusion
coefficients for a given disordered medium. 
Usually, these equations must be numerically solved, except for some simple
cases.
It is showed in this section that for the low temperature limit, some 
analytical predictions may be obtained within the RBM.

\subsection{Self-consistent conditions}

Many authors \cite{alexR,odag81,webm81,summ81} have derived the EMA that
allows to obtain effective diffusion coefficients. 
The approach considers one impure bond of the disordered 
lattice as embedded in an \textit{effective medium}, mimicking the average
surroundings.
By imposing the averaged fluctuations to be zero, the self--consistent 
condition is derived for the transition rate of the effective medium.
For an hypercubic $d$--dimensional isotropic lattice in the long--time 
limit, this condition reads \cite{bouc90}:
\begin{equation}
\label{EMAD}
\left\langle \frac{(\omega -\sigma )}{\omega +(d-1)\sigma }
\right\rangle_{\nu (\omega)}=0,
\end{equation}
where $\omega$ is the transition rate of the impure bond distributed according
to the PDF $\nu (\omega)$, $\sigma $ is the transition rate of the effective 
medium, and the brackets denote average over the PDF $\nu (\omega )$.
Solving the self--consistent condition for $\sigma$, the diffusion
coefficient is obtained as $D=\sigma \ a^{2}$, where $a$ is the lattice
constant.

The anisotropic extension of such formalism, where there exist $n$ different
directions, leads to $n$ coupled equations that self--consistently solve for the $n$ different
diffusion coefficients. In a two--dimensional square lattice, and for the
long--time limit, these conditions are \cite{parr87,reye00}
\begin{equation}
\label{EMAani}
\left\langle \frac{(\omega_m-\sigma_m)}{\omega_m+(f_{mn}^{-1}-1)\sigma_m}
\right\rangle _{\nu_m(\omega_m)}=0,
\end{equation}
with 
\begin{equation}
\label{fij}
f_{mn}=\frac{2}{\pi}\ \arctan \sqrt{\frac{\sigma_m}{\sigma_n}}
\end{equation}
and $m,n=1,2$ denoting the principal axes of the lattice.
The effective transition rates $\sigma _{i}$ are related to the diffusion 
constants by $D_{i}=\sigma _{i} a^2$.

\begin{figure}[!tbp]
\includegraphics[angle=-90,width=8.5cm,clip=true]{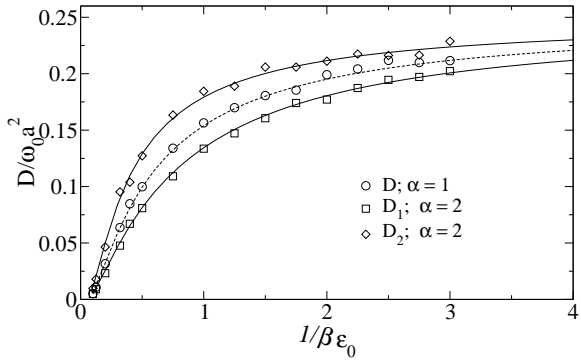}
\caption{\label{f2}
Diffusion coefficients at high temperatures. Lines correspond to the
EMA solution and symbols to numerical simulations. The isotropic
case is represented by a dashed line and the anisotropic case $\alpha = 2$ by
continuous lines.}
\end{figure}

At high temperatures, the particle can easily overcome energy barriers,
and eventually all diffusion constants approach the same value.
In Fig. \ref{f2}, the normalized diffusion coefficients at high
temperatures for the two--dimensional square lattice with an exponential PDF
are plotted as functions of temperature, both under isotropic and anisotropic 
conditions. 
Lines represent the solutions of the EMA self--consistent conditions Eqs. 
(\ref{EMAD}) and (\ref{EMAani}), and symbols correspond to numerical 
simulations (see Sec. \ref{SIV}). The figure shows that, for high temperatures,
$D_i/a^2 \omega_0 \rightarrow 1/z$.
Analogous results are obtained using the uniform PDF.
In the next Subsections the predictions of EMA for diffusion at 
low temperatures are considered.

\subsection{Isotropic case}

For an isotropic hypercubic lattice in $d$ dimensions, Eq. (\ref{EMAD}) 
may be written as 
\begin{equation}
\label{EMAD2}
\left\langle \frac{(\omega +(d-1)\sigma -d\sigma)}
{\omega +(d-1)\sigma}\right\rangle _{\nu (\omega )}=0.
\end{equation}
By introducing the explicit dependence on energy $\omega =\omega (E)$ given in
Eq. (\ref{wdeE}), and transforming the transition rate average over 
$\nu(\omega)$ to an energy average over $\rho (E)$, Eq. (\ref{EMAD2}) becomes 
\begin{equation}
\label{EMAD3}
\left\langle \frac{1}{\omega (E)+(d-1)\sigma}
\right\rangle _{\rho (E)}=\frac{1}{d\sigma }.
\end{equation}
For $\beta \rightarrow \infty$ the transition rate $\omega (E)$
varies extremely rapidly, due to its exponential dependence. 
Therefore it is possible to define an energy
value $E_c$ such that two possibilities arise: 
$\omega (E) \gg (d-1) \sigma $ for $E<E_c$, 
or $\omega (E) \ll (d-1) \sigma$ for $E>E_c$.
The characteristic value $E_c$ can be therefore defined as
\begin{equation}
\label{sdeEd}
\omega (E_c) =(d-1) \sigma .
\end{equation}
For values of $E<E_c$, the left hand side of Eq. (\ref{EMAD3}) vanishes.
Alternatively, for $E>E_c$ the value 
$\omega (E)$ in the left hand side of Eq. (\ref{EMAD3}) may be ignored.
Taking these conditions into account, and averaging over $\rho (E)$,
Eq. (\ref{EMAD3}) for $\beta \rightarrow \infty$ becomes 
\begin{equation}
\frac{1}{d\sigma }=\int\limits_{E_c}^{\infty }\frac{\rho (E) }{(d-1)\sigma }dE.
\end{equation}
Or, equivalently, 
\begin{equation}
\label{CPPdEMA}
\int\limits_{0}^{E_c}\rho (E) dE=\frac{1}{d}.
\end{equation}

In the EMA, the bond percolation threshold of the hypercubic lattice is 
given by $p_c^{EMA}=d^{-1}$ \cite{alexR,odag81,webm81,summ81}.
Therefore, Eq. (\ref{CPPdEMA}) is a condition over $E_c$ for each particular 
PDF $\rho (E)$, in terms of the percolation properties of the lattice. 
Indeed, combining this value of $E_c$ with Eq. (\ref{sdeEd}), the EMA 
diffusion coefficient for isotropic $d$ dimensional hypercubic lattices 
at low temperatures is given by 
\begin{equation}
\label{sdeEcd}
D =\frac{\omega _0 a^2}{z(d-1)}e^{-\beta E_c}.  
\end{equation}
It is worth noting that $p_c^{EMA}=d^{-1}$ is only exact for $d=2$ 
\cite{stauffer}, therefore Eq. (\ref{sdeEcd}) does not give the correct 
exponential behavior for $d=3$, 
and other approximations, such as CPPA, should be considered \cite{amba95}.

\subsection{Anisotropic two--dimensional case}

For the anisotropic two--dimensional case, Eq. (\ref{EMAani}) turns into
two self--consistent conditions,
with transition rate PDFs $\nu_1 (\omega_1)$ and $\nu_2 (\omega_2)$,
for each direction of the lattice.
By introducing the energy dependence $\omega_1=\omega (E_1)$
and $\omega_2=\omega (E_2)$, the two self--consistent conditions read

\begin{eqnarray}
\left\langle 
\frac{1}{\omega \left( E_{1}\right) +\left( f_{12}^{-1}-1\right) \sigma _1}
\right\rangle _{\rho _{1}\left( E_{1}\right) }=\frac{f_{12}}{\sigma _1},
\nonumber \\* 
\left\langle 
\frac{1}{\omega \left( E_{2}\right) +\left( f_{21}^{-1}-1\right) \sigma _2}
\right\rangle _{\rho _{2}\left( E_{2}\right) }=\frac{f_{21}}{\sigma _2}.
\end{eqnarray}

Again, the PDFs change abruptly for $\beta \rightarrow \infty $, and a parameter
$E_c$ can be defined as in the isotropic case.
However, $E_c$ is expected to be characteristic of the
underlying energy landscape, so the diffusion coefficients in each direction
are expected to be governed by a single $E_c$.
For the anisotropic case, an energy $E_c$ will be defined separating two
limiting conditions simultaneously: 
$\omega \left(E_{1}\right) \ll \left( f_{12}^{-1}-1\right) \sigma _{1}$ and 
$\omega \left(E_{2}\right) \ll \left( f_{21}^{-1}-1\right) \sigma _{2}$ 
for $E_1$ and $E_2$ larger than $E_c$, and 
$\omega \left( E_{1}\right) \gg \left( f_{12}^{-1}-1\right) \sigma _{1}$ and 
$\omega \left( E_{2}\right) \gg \left( f_{21}^{-1}-1\right) \sigma _{2}$ 
for $E_1$ and $E_2$ smaller than $E_c$. 
Thus, $E_c$ must verify two simultaneous conditions: 
\begin{eqnarray}
\label{sdeEani}
\omega (E_c) =(f_{12}^{-1}-1)\sigma_1, \nonumber \\*
\omega (E_c) =(f_{21}^{-1}-1)\sigma_2.
\end{eqnarray}
In this way, a set of equations analogous to Eq. (\ref{CPPdEMA}) are obtained,
\begin{eqnarray}
\label{ecani1}
\int\limits_{0}^{E_c}\rho _1(E_1) dE_1=f_{12}, \nonumber \\*
\int\limits_{0}^{E_c}\rho _2(E_2) dE_2=f_{21}.
\end{eqnarray}
By adding Eqs. (\ref{ecani1}), using Eq. (\ref{fij}) and trigonometric
relations, an expression is arrived at,
\begin{equation}
\label{ecani2}
\int\limits_{0}^{E_c}\rho _1(E_1) dE_1+\int\limits_{0}^{E_c}\rho _2(E_2) dE_2=1,
\end{equation}
which gives the activation energy $E_c$ as a function of the anisotropy
$\alpha$.
Moreover, by replacing the expressions in Eqs. (\ref{ecani1}) for $f_{12}$ and
$f_{21}$ in Eqs. (\ref{sdeEani}), and solving for $\sigma _1$ and $\sigma _2$, 
the corresponding anisotropic diffusion coefficients are obtained,
\begin{eqnarray}
\label{sdeEcani}
D_1=\frac{\int\limits_{0}^{E_c}\rho_1(E_1) dE_1}
{\int\limits_{0}^{E_c}\rho_2(E_2) dE_2}
\frac{\omega_0 a^2}{z}e^{-\beta E_c},\nonumber \\*
D_2=\frac{\int\limits_{0}^{E_c}\rho_2(E_2) dE_2}
{\int\limits_{0}^{E_c}\rho_1(E_1) dE_1}
\frac{\omega_0 a^2}{z}e^{-\beta E_c}.
\end{eqnarray}
The isotropic result Eq. (\ref{sdeEcd}) is obviously recovered by setting 
$\rho_1=\rho_2$.

\begin{figure}[!tbp]
\includegraphics[angle=-90,width=8.5cm,clip=true]{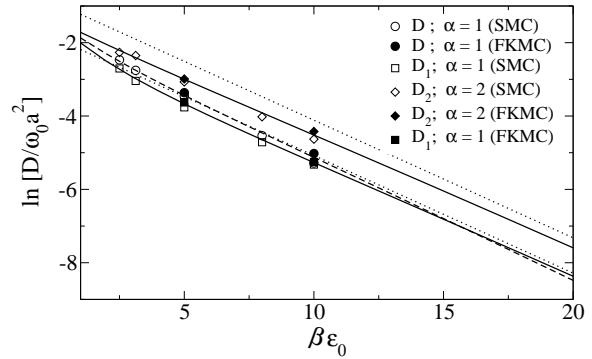}
\caption{\label{f3}
Arrhenius plot of the diffusion coefficients for an exponential PDF. The
solution of the self--consistent EMA conditions is represented with a dashed 
line for the isotropic case, and with continuous lines for the anisotropic 
$\alpha =2$ cases. Dotted lines represent the analytical EMA predictions, 
Eqs. (\ref{sdeEcani}), for low temperatures. Symbols correspond to SMC and 
FKMC simulations, as indicated.}
\end{figure}

\begin{figure}[!tbp]
\includegraphics[angle=-90,width=8.5cm,clip=true]{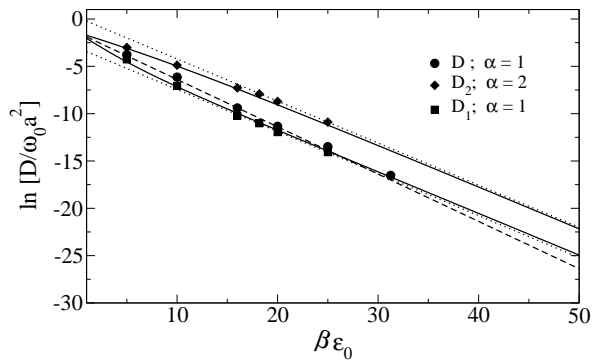}
\caption{\label{f4}
Arrhenius plot of the diffusion coefficients for a uniform PDF. The
solution of the self--consistent EMA conditions is  represented with a dashed 
line for the isotropic case, and with continuous lines for the anisotropic 
$\alpha =2$ cases.
Dotted lines represent the analytical EMA predictions, Eqs. (\ref{sdeEcani}), 
for low temperatures. Symbols correspond to FKMC simulations.}
\end{figure}

Figures \ref{f3} and \ref{f4} show Arrhenius plots of the anisotropic
diffusion coefficients at low temperatures, corresponding to the exponential
and uniform PDFs, respectively.
The solution of EMA self--consistent conditions Eqs. (\ref{EMAD}) and 
(\ref{EMAani}) are represented with dashed an continuous lines, for the 
isotropic and anisotropic $\alpha =2$ cases, respectively.
Dotted lines represent the prediction of EMA for the low temperature limit,
Eqs. (\ref{sdeEcani}). 
Symbols are the results of numerical simulations, as described in the next 
section.
These figures show that the diffusion coefficients in each direction
follow Arrhenius laws with the same activation energy. 
Even though simulation data at lower temperatures are needed, the
agreement with the low temperature anisotropic diffusion coefficients is better
for a uniform PDF than for an exponential PDF.

\section{Numerical simulations\label{SIV}}

Monte Carlo simulations were performed to obtain long--time diffusion 
coefficients for comparison with anisotropic EMA predictions.
The energy landscape was selected from the corresponding PDF at $t=0$ and 
kept fixed during the diffusion process. 
At $t=0$ a particle was assigned to a random initial site $i$.
Different Monte Carlo algorithms may be used at this point and
two possibilities were considered: standard Monte Carlo (SMC) 
\cite{argy95} and a fast kinetic Monte Carlo (FKMC) \cite{buln98} scheme.
A brief description of these methods is given in the following.

In SMC, the particle selects at random one of its nearest neighbors $j$ 
and tries to overcome the barrier between them in a time unit. 
A random number $\xi \in \left( 0,1\right)$ is generated such that if 
$\xi <\omega _{ij}$ the jump is
effective, otherwise the particle stays at the initial site. 
In this process, one unit of time is used for every jump trial.
Although SMC simulations proved to be very useful for studying diffusion
processes, it was shown that it is not too appropriate for studying 
low temperature regimes \cite{buln98,argy95,dyreR}. 
At low temperatures, the transition rates decrease exponentially with 
increasing $\beta $, and the
random number $\xi $ is, mostly, orders of magnitude greater than the
transition rates, making the number of effective jumps (displacements)
very small and the long--time diffusion regime difficult to reach.

In the FKMC \cite{buln98} scheme, consider the particle in 
a site $i$ on a lattice with its $z$ nearest
neighbors $j$ $(j=1,...,z)$. The transition rates from $i$ to $j$
are denoted $\omega_{ij}$. The total transition rate $\omega_i$ from
site $i$ is defined as: 
\begin{equation}
\omega_i=\sum\limits_{j=1}^{z}\omega_{ji}.
\end{equation}
Instead of selecting the neighbor at random, as in SMC, a neighbor $k$ is
selected for an effective jump given that
\begin{equation}
\frac{1}{\omega_i}\sum\limits_{j=1}^{k-1}\omega_{ji}<\xi_1\leq 
\frac{1}{\omega_i}\sum\limits_{j=k}^{z}\omega_{ji},
\end{equation}
where $\xi _1$ is randomly uniformly distributed in $(0,1)$.
The time variable $t$ is then increased in $t'$,
where $t'$ is chosen from an exponential distribution with mean
waiting time $\omega _i^{-1}$. Therefore,
\begin{equation}
t'=-\frac{1}{\omega _{i}}\ln \xi _{2},
\end{equation}
with $\xi _{2}$ randomly uniformly distributed in $(0,1)$.
This procedure is repeated from site $k$ and so on.
In the FKMC algorithm, each jump trial is effective, meaning that
the particle always jumps to one of its neighbors, and the time elapsed in one
jump is accordingly adjusted.
Furthermore, the FKMC algorithm depends on the ratios
$\omega _{ji}/\omega _{i}$ and consequently it
is not as $\beta $ dependent as $\omega _{ji}$ \cite{buln98}.
This algorithm allows to reach larger values
of $\beta $ in simulations of the diffusion process.

Simulations were performed on $300\times 300$ and $500\times 500$ sites square
lattices for the SMC and FKMC, respectively, with periodic boundary 
conditions.
For each algorithm, the mean square displacements on each direction
$\left\langle r_{1}^{2}(t)\right\rangle $ and 
$\left\langle r_{2}^{2}(t)\right\rangle $ were computed, averaging over
between 2000 and 5000 realizations of the random walks.
The long--time diffusion coefficients were
defined through $\left\langle r_{1(2)}^2(t)\right\rangle =2D_{1(2)}t$, 
and were obtained from the best linear fits to the
long--time mean square displacements.

In Figs. \ref{f3} and \ref{f4}, numerical simulations and EMA results are 
presented together. 
In Fig. \ref{f3}, SMC simulations are plotted up to 
$\beta \epsilon_0 = 10$ and some FKMC simulation points are shown for
comparison. Both algorithms coincide within the numerical precision. 
In Fig. \ref{f4}, only FKMC results are presented up to a value 
$\beta \epsilon_0 = 30$.
Monte Carlo simulations do not completely reach the asymptotic low temperatures 
behavior. However, numerical simulations and EMA seem to agree very well 
in the accessible temperature range. 

\section{Critical percolation path approximation\label{SV}}

The idea of a percolation path governing diffusion at low temperatures was
first developed in Ref. \cite{ambe71} and rigorously proved later
\cite{tyc89,ledo89}.
In this section, this idea will be briefly summarized and extended 
to anisotropic conditions.

At low temperatures the characteristic Arrhenius diffusion energy $E_c$ 
can be related to the bond percolation threshold of the lattice. 
Consider a random walk on a realization of the
disorder energy landscape at a very low temperature.
In order to overcome a barrier with an energy $E'$, the particle spends a mean
waiting time $t' \sim \exp (\beta E')$.
For short times, therefore, the particle can only move to sites which are
connected by low energy barriers and is surrounded by a perimeter of higher
energy.
Roughly, at time $t'$ the
particle might jump barriers with $E\leq E'$, and the probability
to overcome this barriers is $\int_{0}^{E'}\rho (E) dE$.
For longer times, the particle could overcome the lowest barrier
of the perimeter, and access a new region with a higher energy perimeter. 
These regions are non--compact in the sense that they
may have inside barriers that belong to the perimeter barriers. 
Eventually, there exists a particular barrier of height $E_c$, beyond which 
the particle gains accesses to the whole
system, through the corresponding percolation path of energies $E\leq E_c$.
Thus, for an isotropic medium, $E_c$ is given by
\begin{equation}
\label{CPPiso}
\int\limits_{0}^{E_c}\rho (E) dE=p_c,  
\end{equation}
where $p_c$ is the bond percolation threshold of the system ($p_c=0.5$
for the two dimensional isotropic square lattice \cite{stauffer}).
It has been shown that $E_c$ is the highest energy barrier which the particle 
must overcome in order to gain full access to the percolation network.
The long--time diffusion coefficient must therefore be
$D\sim \exp (-\beta E_c)$, which is indeed the observed behavior of
isotropic diffusion at low temperatures \cite{dyre94,argy95}.

The percolation threshold of a particular lattice, which is given by a point 
$p_{c}$ for isotropic percolation, becomes for anisotropic percolation a 
critical surface $\varphi (\{ p_i\} )=0$ \cite{grimmet}, where $\{ p_i\}$ 
denotes the set of relevant occupation probabilities. 
For example, the percolation function is: $\varphi (p)=p-p_c$ for isotropic
percolation, $\varphi (p_1,p_2)=p_1+p_2-1$ for the square lattice, and 
$\varphi (p_1,p_2,p_3) =p_1+p_2+p_3-p_1 p_2 p_3-1$
for the triangular lattice \cite{grimmet}.
Furthermore, the critical surface implies a change in the morphology of the
incipient percolation network.

In the anisotropic RBM context, the occupation probabilities of a
bond with energy barrier $E$, i.e. accessibility condition of the bond,
is given by the probability of $E$ being lower than the maximum accessible
barrier. Therefore, the generalization of Eq. (\ref{CPPiso}) to anisotropic
conditions becomes
\begin{equation}
\label{CPPanigen}
\varphi \left( \left\{ \int\limits_{0}^{E_c}\rho _{i}\left( E_{i}\right)
dE_{i}\right\} \right) =0.
\end{equation}
Note that there exists just one energy $E_c$, which is the same for all 
directions, and gives full access to the whole anisotropic percolation network.
For the anisotropic RBM on a square lattice, Eq. (\ref{CPPanigen}) 
becomes equal to the EMA prediction, Eq. (\ref{ecani2}). 
This means that EMA predicts the correct critical percolation surface 
$\varphi (p_1,p_2)=p_1+p_2-1$ for anisotropic bond percolation in the square
lattice \cite{bern74,tole92}.

\begin{figure}[!tbp]
\includegraphics[angle=-90,width=8.5cm,clip=true]{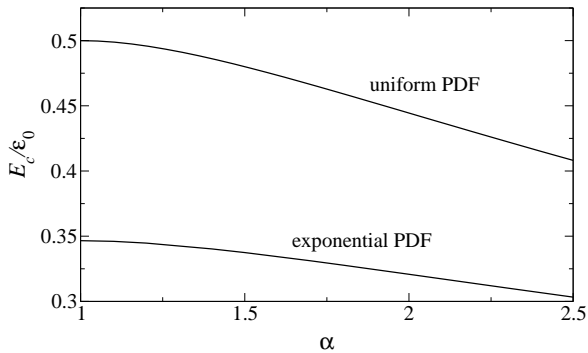}
\caption{\label{f5}
Dependence of the activation energy $E_c$ on the anisotropic parameter $\alpha$
for the exponential and uniform PDFs studied here.}
\end{figure}

Figure \ref{f5} shows the effect of anisotropy $\alpha$ on $E_c(\alpha)$ for 
the energy distributions studied in the present model, and predicted both 
by CPPA Eq. (\ref{CPPanigen}), and EMA Eq. (\ref{ecani2}).
For the exponential PDF, the condition for $E_c$ reads 
$\exp (-E_c/\epsilon_1)+\exp (-E_c/\epsilon_2)=1$,
which was numerically solved.
For the uniform PDF Eq. (\ref{ecani2}) gives 
$E_c/\epsilon_0=2 \alpha/(\alpha+1)^2$.

\section{Concluding remarks\label{SVI}}

In this paper, diffusion properties were studied using an anisotropic RBM, with
emphasis on the low temperature behavior and on percolation properties.
Two kind of PDFs were used to characterize different directions of the lattice,
namely, exponential and uniform PDFs.
The anisotropic EMA was used to calculate the long--time diffusion properties
for all temperatures, derived from the numerical solutions of the 
self--consistent conditions expressed in Eqs. (\ref{EMAani}). 
Furthermore, analytical expressions for
the diffusion coefficients at low temperatures were obtained,
Eq. (\ref{sdeEcani}), which show
that diffusion in different directions follows Arrhenius laws with a same 
activation energy $E_c$. This should be compared with the thermally 
activated diffusion in anisotropic bond percolation lattices, in which 
different activation energies are found for each direction \cite{bust03}. 
In the present model, only one activation energy is found due to the existence
of an anisotropic percolation path of low energy barriers, which governs 
the diffusion process.
Besides of giving the activation energy for diffusion, EMA
predicts the exponential prefactor for diffusion and it dependence with the
anisotropy of the disordered system.

The two Monte Carlo algorithms used here, namely, SMC and FKMC, show
a very good agreement with the EMA predictions for the diffusion coefficients 
in the accessible temperature range.
For a more extensive comparison with EMA, other algorithms should be used.

In the present paper, a connection was established between EMA and CPPA 
ideas, and EMA was shown to predict the correct activation energy for 
anisotropic diffusion in the square lattice. 
For other geometries and dimensions, it is expected that EMA will still 
predict an Arrhenius behavior, but with an activation energy that differs from
that predicted by CPPA. This difference is due to the fact that CPPA uses the
percolation threshold as a parameter, while EMA predicts its own percolation
threshold.
However, EMA is known to predict the correct percolation threshold only
for the square lattice, even in anisotropic conditions \cite{bern74,tole92}.
Concerning CPPA, corrections of the form $\beta^y$ become relevant for
dimensions grater than two, but it is not clear which of both approximations,
EMA or CPPA, give better results \cite{amba95} and a 
systematic comparison turns necessary.
Additional work on this direction is now under progress.

\section{Acknowledgments}

The author want to thanks G. L. Insua for useful discussions.
This work was financially supported by \mbox{CONICET}, Argentina.

\bibliography{barbm}

\end{document}